\documentclass[a4paper,aps,12pt,amsmath,amssymb]{revtex4}
\usepackage[dvips]{graphicx}
\usepackage{threeparttable}
\usepackage[yyyymmdd,hhmmss]{datetime}
\usepackage{subfigure}
\usepackage{hyperref}
\usepackage{indentfirst}
\usepackage{color}
\usepackage{rotating}
\usepackage{multirow}
\begin{document}
%\setpagewiselinenumbers
%\modulolinenumbers[1]
%\linenumbers

% -------------------------------------------------------------------- %
\title{How Molecular Chiralities of Bis(mandelato)borate Anions affect Their Binding Structures with Alkali Metal Ions and Microstructural Properties in Tetraalkylphosphonium Ionic Liquids}
\author{Han-Wen Pei$^a$, Bin Li$^b$, Aatto Laaksonen$^{a,c,d}$, Yong-Lei Wang$^a$$\footnote{Author to whom correspondence should be addressed. E-mail: wangyonl@gmail.com; yonglei.wang@mmk.su.se}$}
\affiliation{$^a$Department of Materials and Environmental Chemistry, Arrhenius Laboratory, Stockholm University, SE-106 91 Stockholm, Sweden\\
$^b$School of Chemical Engineering and Technology, Sun Yat-sen University, Zhuhai 519082, P. R. China\\
$^c$State Key Laboratory of Materials-Oriented and Chemical Engineering, Nanjing Tech University, Nanjing 210009, P. R. China\\
$^d$Centre of Advanced Research in Bionanoconjugates and Biopolymers, Petru Poni Institute of Macromolecular Chemistry Aleea Grigore Ghica-Voda, 41A, 700487 Iasi, Romania
}
\date{\today}

% -------------------------------------------------------------------- %
\begin{abstract}
Spiroborate anions based inorganic electrolytes and ionic liquids (ILs) have fascinating electrochemical and tribological properties, and have received widespread attention in industrial applications.
Molecular chiralities of spiroborate anions have a significant effect on microstructures and macroscopic functionalities of these ionic materials in applications, and thus deserve a fundamental understanding.
In current work, we performed quantum chemistry calculations to address binding strength and coordination structures of chiral bis(mandelato)borate ([BMB]) anions with representative alkali metal ions, as well as electronic properties of alkali metal ion-[BMB] ion pair complexes.
The optimized [BMB] conformers are categorized into V-shaped, bent, and twisted structures with varied electrostatic potential contours, conformational energies, and distinct alkali metal ion-[BMB] binding structures.
Alkali metal ions have additional associations with phenyl groups in V-shaped [BMB] conformers owing to preferential cation-$\pi$ interactions.
Furthermore, effects of molecular chiralities of [BMB] anions on thermodynamics and microstructural properties of tetraalkylphosphonium [BMB] ILs were studied by performing extensive atomistic interactions.
Oxygen atoms in [BMB] anions have competitive hydrogen bonding interactions with hydrogen atoms in cations depending on molecular chiralities and steric hindrance effects of [BMB] anions.
However, molecular chiralities of [BMB] anions have negligible effect on liquid densities of tetraalkylphosphonium [BMB] ILs and spatial distributions of boron atoms in anions around phosphorous atoms in cations.
Enlarging tetraalkylphosphonium cation sizes leads to enhanced cation-anion intermolecular hydrogen bonding and Coulombic interactions due to enhanced segregation of polar groups in apolar networks in heterogeneous IL matrices as verified from scattering structural functions.
\end{abstract}
\maketitle

% -------------------------------------------------------------------- %
\clearpage
\section{Introduction}

\par
Ionic liquids (ILs) represent an intriguing category of molten salts solely composed of inorganic or organic anions and, most commonly, organic cations having melting points at or close to room temperature~\cite{castner2011ionic, hayes2015structure, zhang2016catalytic, bedrov2019molecular}.
In recent years, ILs have received significant attention in diverse academia and industrial communities due to their multifaceted physicochemical properties, such as non-flammability, negligible volatility, reasonable viscosity-temperature features, high thermal-oxidative stabilities, wide electrochemical windows, as well as outstanding affinities to polar and apolar compounds~\cite{armand2009ionic, zhou2009ionic, castner2011ionic, wang2012ionic, hayes2015structure, dai2017ionic, bedrov2019molecular}.
These fascinating characteristics can be widely tuned in a controllable fashion via a judicious selection of ion moieties and by mutating specific atoms in constituent ion species~\cite{kashyap2013does, wang2014atomistic, wu2016structure, jankowski2016new, wang2017microstructures}, and thus render ILs exceptionally attractive and reliable alternatives to conventional molecular liquids in applications spanning from solvents in materials synthesis to electrolytes in electrochemical devices~\cite{armand2009ionic, zhou2009ionic, castner2011ionic, abbott2013electroplating, zhang2016catalytic, dai2017ionic, watanabe2017application, bedrov2019molecular}.

\par
ILs consisting of spiroborate anions coupled with tetraalkylphosphonium, pyrrolidinium, and imidazolium cations present practical tribological advantages, and can be used as alternative high-performance lubricants and lubricant additives in tribology due to their outstanding friction reducing and anti-wear performance in comparison with conventional fully formulated engine oil in tribological contacts with a wide variety of solid materials~\cite{shah2011novel, shah2013boron, taher2014halogen, gusain2017physicochemical, an2018friction, pilkington2018electro, shah2018interfacial, rohlmann2019non, rutland2019langmuir}.
Therefore, the number of fundamental studies and industrial applications of spiroborate anions based ILs is growing rapidly in recent years.

\par
Besides variations in cation structures, molecular chiralities of spiroborate anions have a significant effect on microstructures and dynamics, mesoscopic liquid morphologies, and macroscopic functionalities of these ILs in tribology, electrochemistry, and pharmaceutical chemistry~\cite{yu2008chiral, absalan2012determination, sedghamiz2019chiral}.
The bis(mandelato)borate ([BMB]) anion based ILs are used as selectors for chiral discrimination of propranolol enantiomers~\cite{sedghamiz2018chiral, sedghamiz2019chiral}.
Both experimental~\cite{absalan2012determination} and computational studies~\cite{sedghamiz2018chiral, sedghamiz2019chiral} showed that [BMB] anions contribute to distinct hydrogen bonding (HB) and $\pi$-$\pi$ stacking interactions with propranolol enantiomers, leading to the formation of propranolol-[BMB] complexes with varied molecular stabilities depending on specific chiralities of [BMB] anions.
In another work, Wong and coworkers presented experimental evidences that [BMB] anions are effective resolving agents for resolution of a chemically diverse range of racemic cations via metathesis crystallization~\cite{wong2015mandelato, wong2017chiral, wong2018isolation}.

\par
\begin{figure}[!h]
\centering\includegraphics[width=0.5\textwidth]{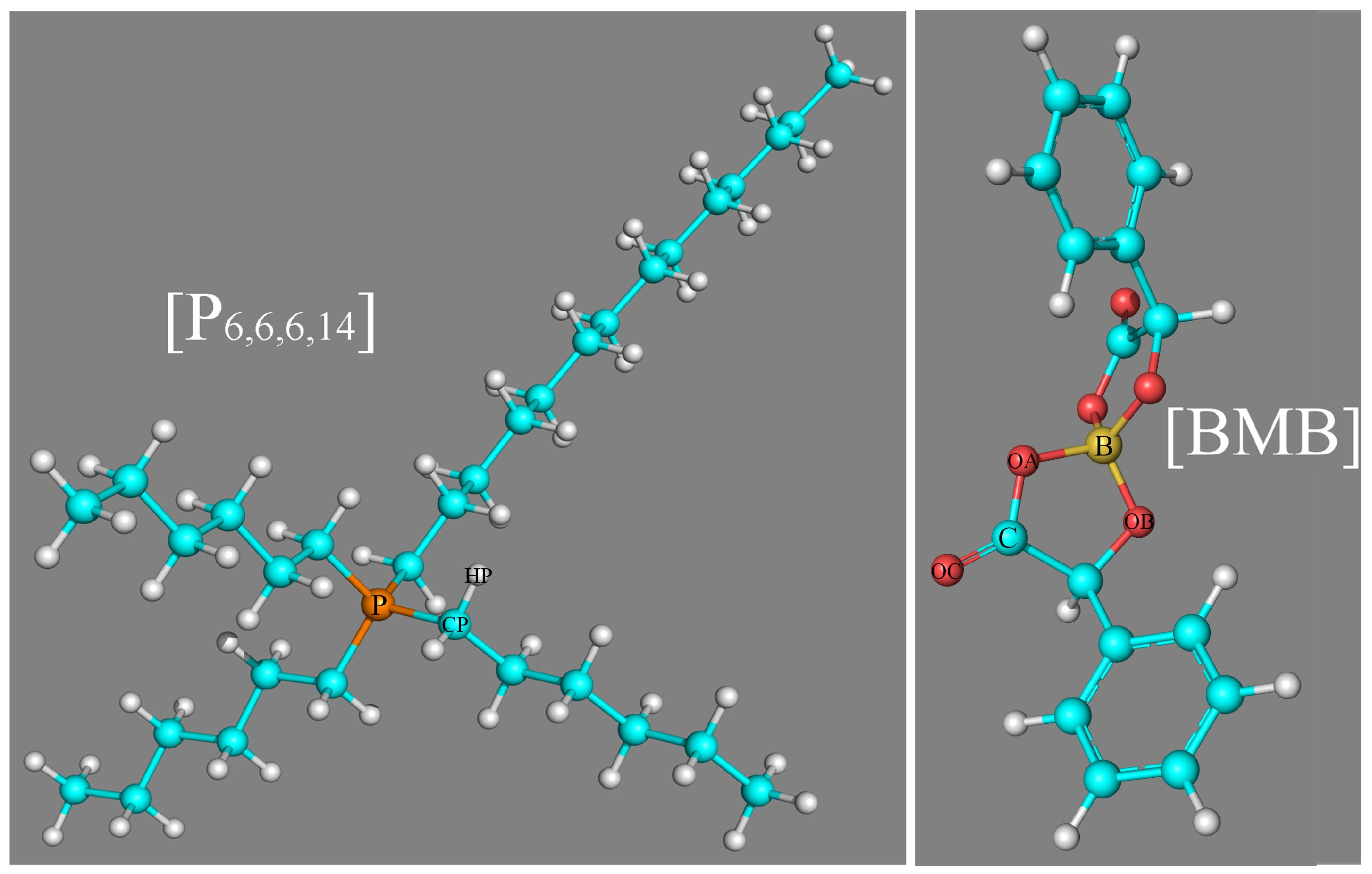}
\caption{Molecular structures of [P$_{\textrm{6,6,6,14}}$] cation and B(S)-Man(SR) [BMB] conformer, as well as representative atom types in these two ion species.}\label{fig:il_structure}
\end{figure}

The significance of [BMB] anions with different molecular chiralities being sensitive to local ionic environments motivates us to explore their delicate associations with alkali metal ions and tetraalkylphosphonium cations, which may provide valuable information for selection and design of suitable electrolytes and lubricant additives with desirable physicochemical, structural, and functional properties for their applications in electrochemistry and tribology~\cite{shah2011novel, shah2013boron, jankowski2016new, an2018friction, pilkington2018electro, shah2018interfacial,  rutland2019langmuir, franco2019boosting}.
In current work, we performed intensive density functional theory (DFT) calculations to study effects of molecular chiralities of [BMB] anions on their specific binding structures with representative alkali metal ions, and thereafter extensive atomistic simulations to explore delicate interactions of [BMB] anions with tetraalkylphosphonium cations having varied alkyl substituents.
Three tetraalkylphosphonium cations, including tetrabutylphosphonium ([P$_{\textrm{4,4,4,4}}$]), tributyloctylphosphonium ([P$_{\textrm{4,4,4,8}}$]) and trihexyltetradecylphosphonium ([P$_{\textrm{6,6,6,14}}$]), and six [BMB] conformers having different molecular chiralities are considered in present work.
A schematic molecular structure of [P$_{\textrm{6,6,6,14}}$] cation and B(S)-Man(SR) [BMB] conformer, as well as representative atom types in these two ions are present in Fig.~\ref{fig:il_structure}.

% -------------------------------------------------------------------- %
\section{Binding Structures of [BMB] Anions with Alkali Metal Ions}

\par
DFT calculations were first performed to optimize molecular geometries of single [BMB] conformers, and thereafter their tight binding ion pair structures with alkali metal ions, respectively, using Gaussian 09 package~\cite{g09} (version D.01) at B3LYP/6-311+G(d) level of theory~\cite{lee1988development} with Grimme’s-D3 dispersion correction~\cite{grimme2010consistent}.
This dispersion correction is appropriate over medium ($\approx$ 2-5 $\textrm{\AA}$) and long ranges ($>$ 5 $\textrm{\AA}$), and is an effective method to obtain binding structures of complexes with reduced computational cost.
Only molecular geometries with all positive vibrational frequencies were taken into consideration to ensure that the obtained anion structures of [BMB] conformers are in fact true minima in the corresponding energy landscapes.

\begin{figure}[!h]
\centering\includegraphics[width=0.8\textwidth]{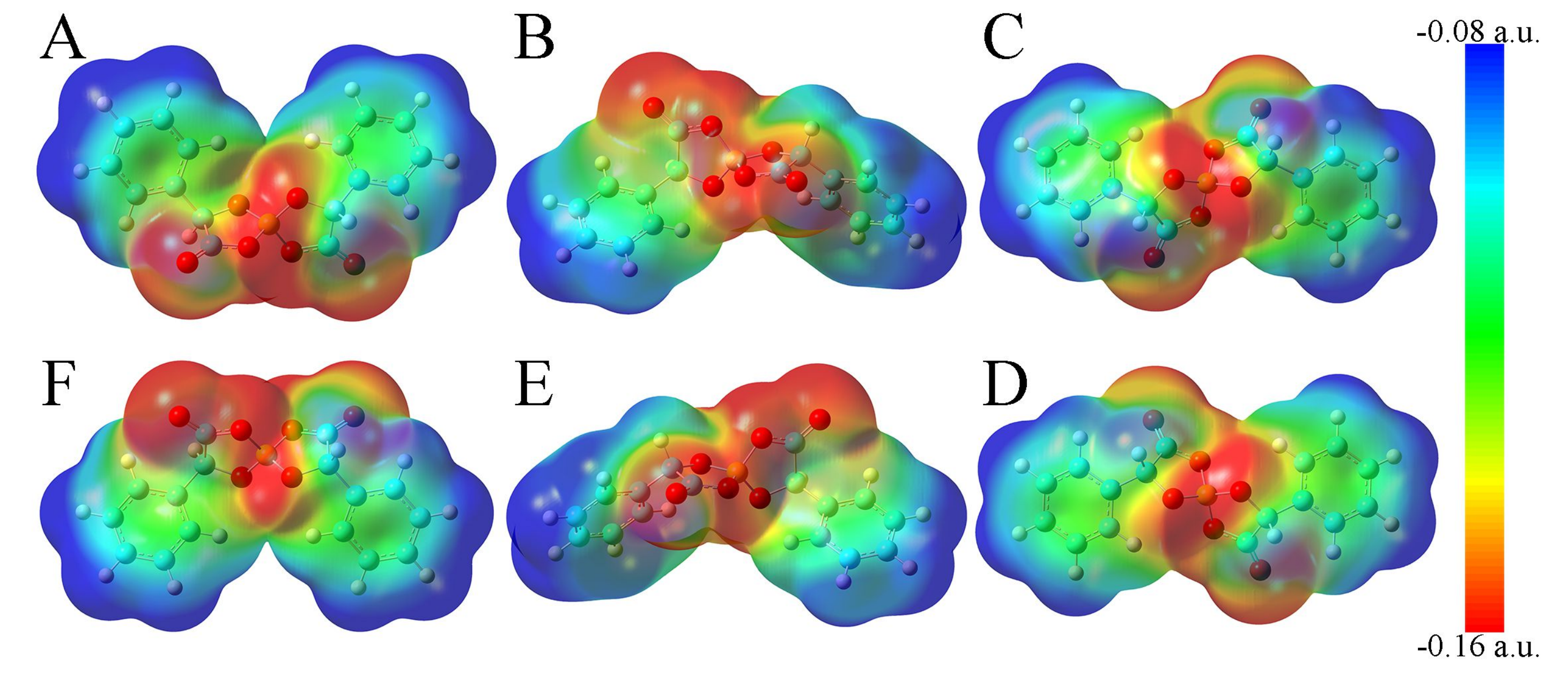}
\caption{Molecular electrostatic potential contours of six [BMB] conformers with different molecular chiralities obtained from DFT calculations. (A) B(R)-Man(RR), (B) B(R)-Man(RS), (C) B(R)-Man(SS), (D) B(S)-Man(RR), (E) B(S)-Man(SR), and (F) B(S)-Man(SS).}\label{fig:bmb_single}
\end{figure}

\par
The molecular electrostatic potential contours of six [BMB] conformers based on their respective optimized ion structures are present in Fig.~\ref{fig:bmb_single}.
These optimized [BMB] conformers have remarkable molecular shapes, and can be categorized into three pairs, that is, the B(R)-Man(RR) and B(S)-Man(SS) [BMB] conformers (panels A and F in Fig.~\ref{fig:bmb_single}) are characterized by V-shaped structures, the B(R)-Man(RS) and B(S)-Man(SR) [BMB] conformers (panels B and E in Fig.~\ref{fig:bmb_single}) are described by bent conformations, and the B(R)-Man(SS) and B(S)-Man(RR) [BMB] conformers (panels C and D in Fig.~\ref{fig:bmb_single}) have twisted configurations, respectively.
Each pair has symmetric ion structures and electrostatic potential contours, and is characterized with similar relative conformational energies, dipole moments, and molecular volumes.
These computational data are listed in Table~\ref{tbl:bmb_data}.
Previous DFT calculations indicated that boron-based chiral anions are energetically similar in diastereomeric [BMB] systems~\cite{wong2018isolation}.
In current work, the B(R)-Man(RS) and B(S)-Man(SR) [BMB] conformers have the lowest conformational energies, whereas the B(R)-Man(SS) and B(S)-Man(RR) [BMB] conformers have the highest conformational energies, respectively, attributing to unbalanced intramolecular forces within [BMB] anionic frameworks.
It is noted that the conformational energy difference between all these [BMB] conformers is less than 1 kJ/mol, indicating that these [BMB] conformers may co-exist in IL samples, which is consistent with that obtained from NMR spectra~\cite{wong2015mandelato}.
However, the precipitation of twisted B(R)-Man(SS) and B(S)-Man(RR) [BMB] conformers is always observed in crystallized salts, rather than the V-shaped B(R)-Man(RR) and B(S)-Man(SS) [BMB] conformers~\cite{wong2015mandelato, wong2017chiral}.
DFT calculations showed that in all [BMB] conformers four oxygen atoms exhibit tetrahedral-like distributions around central boron atoms, whereas two oxalato rings have constrained distributions.
The addition of phenyl groups to central oxalato moieties results in distorted orientations of phenyl groups around oxalato rings.
These distinct conformational variations of [BMB] anions lead to their varied dipole moments and molecular volumes as listed in Table~\ref{tbl:bmb_data}.
The central moieties in [BMB] anions are polar and have significant negative charge, leading to varied HB capabilities of three oxygen atom types with HP atoms (as labelled in Fig.~\ref{fig:il_structure}) in tetraalkylphosphonium cations~\cite{wang2015atomistic}, and with polar solute molecules, like water~\cite{wang2015atomistic, wang2016solvation}.

\begin{table}[!h]
\centering\caption{Relative conformational energies (RCEs) (in kJ/mol), dipole moments (DMs) (in Debye), and molecular volumes (in cm$^3$/mol) of single [BMB] conformers, and relative binding energies (RBEs) (in kJ/mol) of varied [BMB] conformers in coordinating alkali metal ions (Li$^+$, Na$^+$, and K$^+$) determined from DFT calculations.}\label{tbl:bmb_data}
\begin{tabular}{l@{\quad}c@{\quad}c@{\quad}c@{\quad}c@{\quad}r@{\quad}r@{\quad}r}
\hline\hline
&\multicolumn{3}{c}{Single [BMB] conformers} & &\multicolumn{3}{c}{Alkali metal ion-[BMB] ion pairs}\\
\cline{2-4}\cline{6-8}
Conformers       & RCEs   & DMs    & Volumes  & & Li-BMB  & Na-BMB  & K-BMB   \\
\hline
B(R)-Man(RR) & 0.0982 & 6.8945 & 191.770 & & -3.0179 & -7.5136 & -9.5608 \\
B(R)-Man(RS) & 0.0000 & 4.9898 & 225.871 & & -0.3833 & -1.1197 & -4.2263 \\
B(R)-Man(SS) & 0.4638 & 1.6472 & 248.534 & & -0.1394 & -0.8400 &  0.0000 \\
B(S)-Man(RR) & 0.4642 & 1.6474 & 249.508 & &  0.0000 &  0.0000 & -0.3294 \\
B(S)-Man(SR) & 0.0020 & 4.9898 & 226.122 & & -1.9355 & -0.9143 & -4.6534 \\
B(S)-Man(SS) & 0.0986 & 6.8947 & 192.447 & & -3.0175 & -7.5135 & -8.9812 \\
\hline\hline
\end{tabular}
\end{table}

\par
\begin{figure}[!h]
\centering\includegraphics[width=0.8\textwidth]{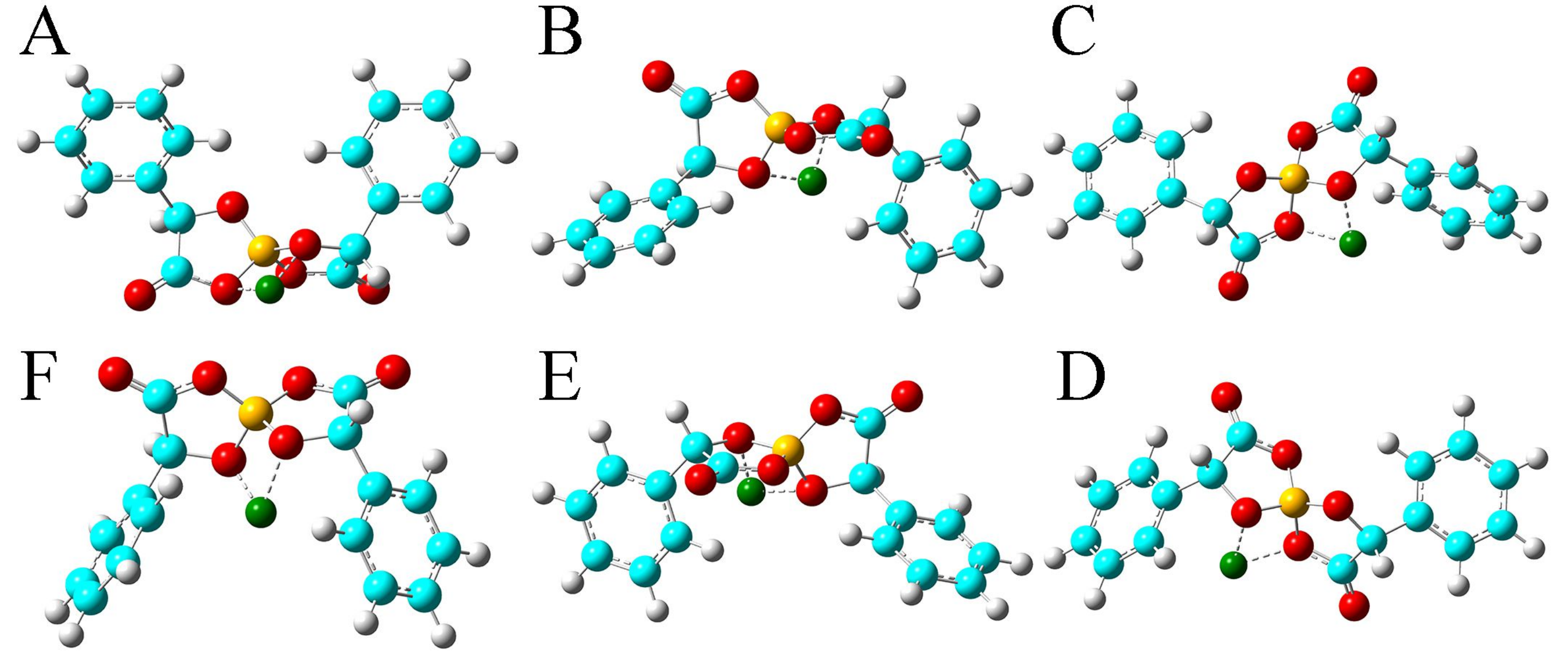}
\caption{Optimized ion pair structures of Li$^+$ ion in coordinating [BMB] anions having varied molecular chiralities determined from DFT calculations. The hydrogen, boron, carbon, and oxygen atoms in [BMB] anions are colored in white, yellow, cyan, and red beads, and Li$^+$ ions are represented by green spheres, respectively. (A) B(R)-Man(RR), (B) B(R)-Man(RS), (C) B(R)-Man(SS), (D) B(S)-Man(RR), (E) B(S)-Man(SR), and (F) B(S)-Man(SS).}\label{fig:bmb_li}
\end{figure}
\begin{figure}[!h]
\centering\includegraphics[width=0.8\textwidth]{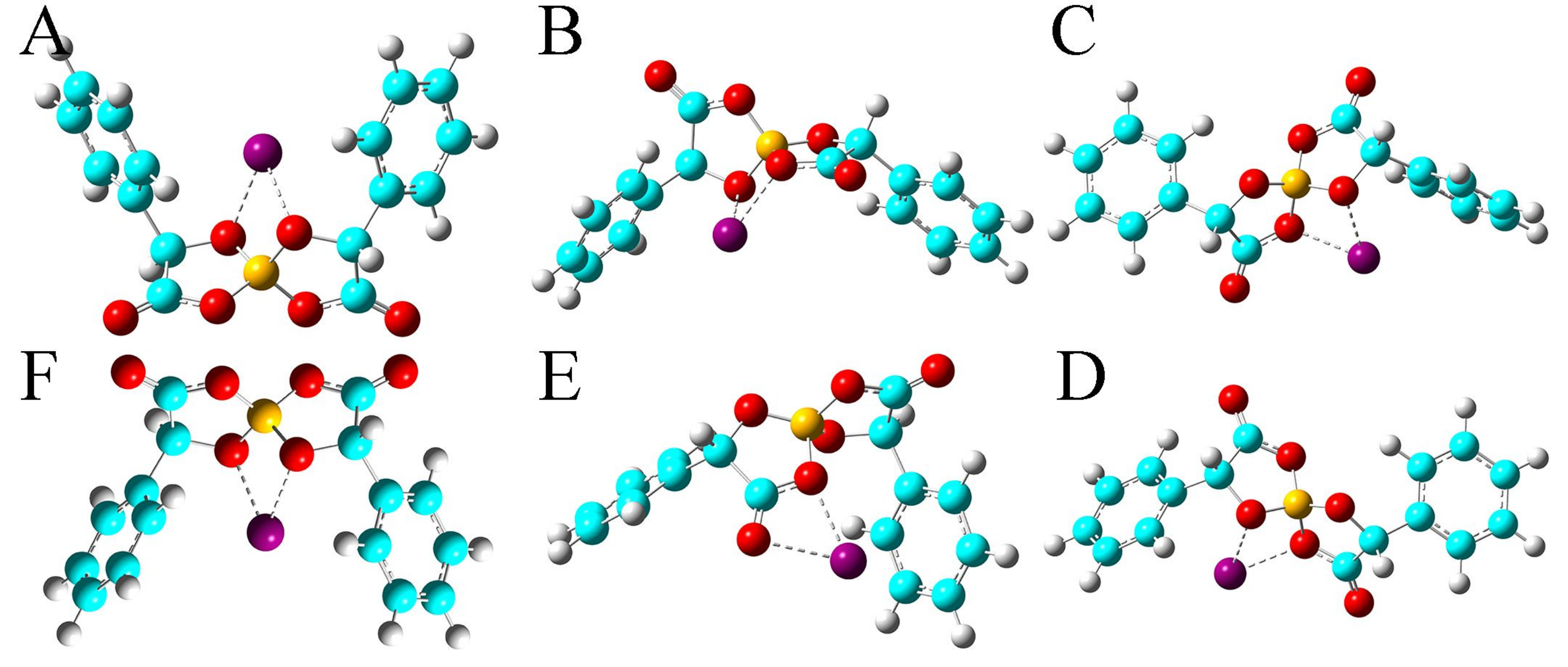}
\caption{Optimized ion pair structures of Na$^+$ ion in coordinating [BMB] anions having varied molecular chiralities determined from DFT calculations. The hydrogen, boron, carbon, and oxygen atoms in [BMB] anions are colored in white, yellow, cyan, and red beads, and Na$^+$ ions are represented by purple spheres, respectively. (A) B(R)-Man(RR), (B) B(R)-Man(RS), (C) B(R)-Man(SS), (D) B(S)-Man(RR), (E) B(S)-Man(SR), and (F) B(S)-Man(SS).}\label{fig:bmb_na}
\end{figure}
\begin{figure}[!h]
\centering\includegraphics[width=0.8\textwidth]{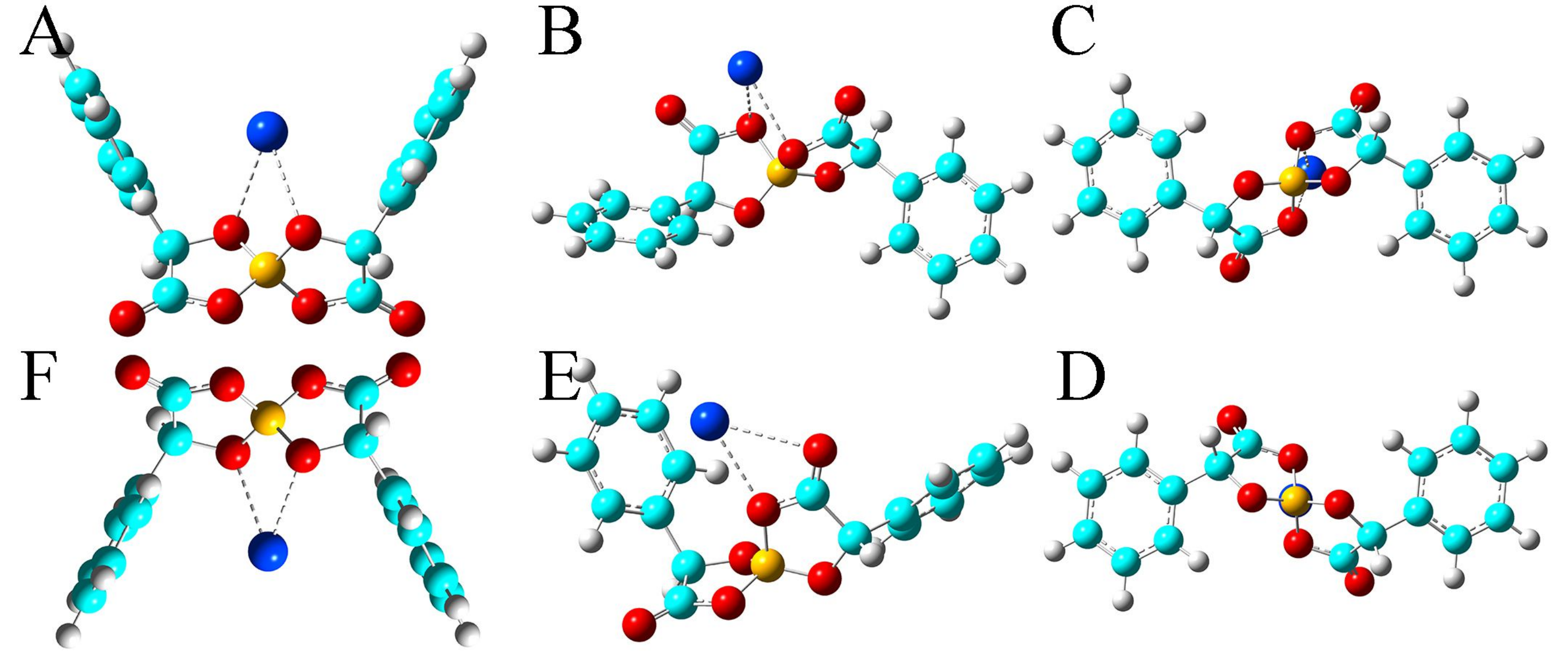}
\caption{Optimized ion pair structures of K$^+$ ion in coordinating [BMB] anions having varied molecular chiralities determined from DFT calculations. The hydrogen, boron, carbon, and oxygen atoms in [BMB] anions are colored in white, yellow, cyan, and red beads, and K$^+$ ions are represented by blue spheres, respectively. (A) B(R)-Man(RR), (B) B(R)-Man(RS), (C) B(R)-Man(SS), (D) B(S)-Man(RR), (E) B(S)-Man(SR), and (F) B(S)-Man(SS).}\label{fig:bmb_kk}
\end{figure}

In addition, the optimized binding structures of varied [BMB] conformers with representative alkali metal ions (Li$^+$, Na$^+$, and K$^+$) were determined from intensive DFT calculations at the same level of theory as that used for the calculation of single [BMB] conformers.
Furthermore, the binding energies (in kJ/mol) of varied [BMB] conformers with alkali metal ions were determined as the energy difference between the optimized alkali metal ion-[BMB] complexes and the summation of conformational energies of separately optimized ion species.
These binding energies are corrected for basis sets superposition error (BSSE) using counterpoise procedure~\cite{boys1970calculation}.
The relative binding energies of three alkali metal ions with six [BMB] conformers are listed in Table~\ref{tbl:bmb_data}, and the corresponding optimized ion pair structures are shown in Fig.~\ref{fig:bmb_li}, Fig.~\ref{fig:bmb_na}, and Fig.~\ref{fig:bmb_kk}, respectively.

\par
All these alkali metal ions exhibit specific associations with central polar moieties of [BMB] conformers via coordinating oxygen atoms in [BMB] anions.
For a given alkali metal ion, both B(R)-Man(RR) and B(S)-Man(SS) [BMB] conformers have the most stable alkali metal ion-[BMB] binding structures, as verified from the relative binding energies shown in Table~\ref{tbl:bmb_data}, due to the intrinsic V-shaped distributions of two phenyl groups in [BMB] anionic frameworks.
All three alkali metal ions are intrinsically coordinated with B(R)-Man(RR) and B(S)-Man(SS) [BMB] conformers in local cavities of V-shaped conformations.
For bent and twisted [BMB] conformers, alkali metal ions either coordinate with two oxygen atoms in central BO$_4$ moieties, or with one oxygen in BO$_4$ moieties and the other one in C=O moieties depending on the delicate interactions of alkali metal ions with specific atoms in [BMB] conformers.

\par
In addition, alkali metal ions have preferential associations with phenyl groups in V-shaped B(R)-Man(RR) and B(S)-Man(SS) [BMB] conformers.
It is demonstrated that an increase of alkali metal ion sizes from Li$^+$ to Na$^+$ and K$^+$ ions leads to their distinct cation-$\pi$ coordinations with phenyl groups~\cite{dougherty2012cation}.
Larger alkali metal ions have stronger cation-$\pi$ interactions and therefore have larger binding energies with V-shaped B(R)-Man(RR) and B(S)-Man(SS) [BMB] conformers.
This may lead to distinct applications of [BMB] anions in alkali metal ion batteries, in which alkali metal ion-[BMB] adducts are effective charge carriers contributing to significant ion conductivities in electrochemical devices.

\begin{figure}[!h]
\centering\includegraphics[width=0.5\textwidth]{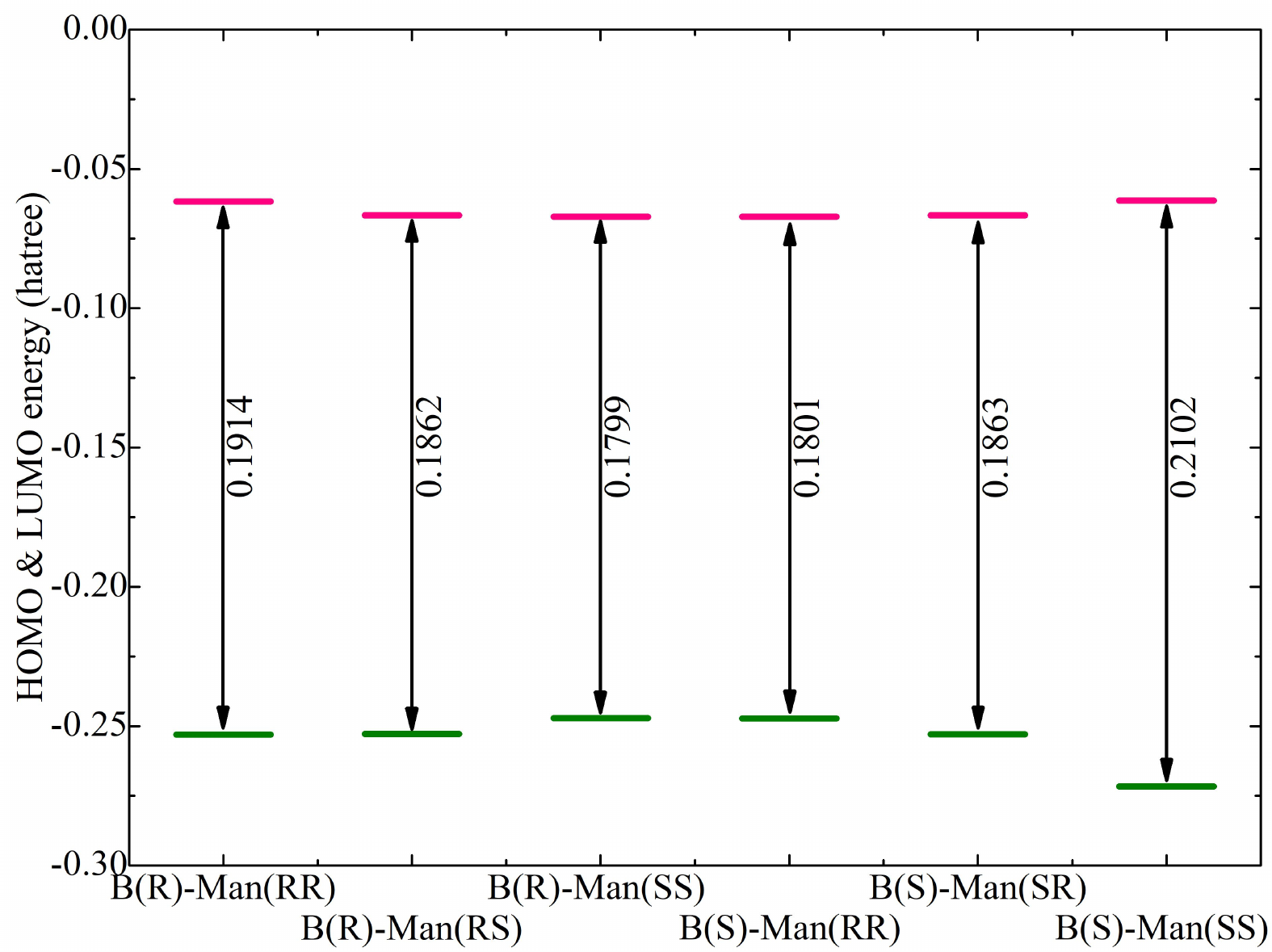}
\caption{Computational HOMO and LUMO energies (in Hartree), and energy gaps (LUMO-HOMO) for Li[BMB] ion pair complexes with anions having varied molecular chiralities. These computational results were obtained from DFT calculations at B3LYP/6-311+G(d) level of theory with Grimme’s-D3 dispersion correction.}\label{fig:libmb_homo_lumo_ene}
\end{figure}

\par
Additional electronic properties of alkali metal ion-[BMB] complexes were calculated based on their optimized ion pair structures.
The energy levels of the highest occupied molecular orbital (HOMO) and the lowest unoccupied molecular orbital (LUMO) for the optimized Li[BMB] ion pair structures are shown in Fig.~\ref{fig:libmb_homo_lumo_ene}.
According to the frontier molecular orbital theory, the HOMO and LUMO energies are useful for evaluating electron donating and accepting abilities of ion species~\cite{dong2012understanding, garcia2015adsorption}.
Higher values of HOMO energies and lower values of LUMO energies indicate a larger tendency for ion species to donate and accept electrons, respectively.
In addition, the HOMO and LUMO energies are directly related to ionization potential and electron affinity, and give quantitative information on the possible electron transition between alkali metal ions and [BMB] anions.
The energy difference between HOMO and LUMO is an important parameter determining molecular-electrical transport properties, molecular stabilities, and liquid phase electrochemical windows of alkali metal ion-[BMB] complexes~\cite{dong2012understanding, garcia2015adsorption, yllo2019experimental}.
It is shown that the HOMO and LUMO energy levels and energy gaps for six Li[BMB] ion pair complexes generally follow a similar tendency as the relative binding energies (listed in Table~\ref{tbl:bmb_data}) for these [BMB] conformers in coordinating Li$^+$ ions.
The twisted B(R)-Man(SS) and B(S)-Man(RR) [BMB] conformers have the lowest energy gap among three categories, indicating that Li$^+$ ions tend to have higher reactivities with twisted [BMB] configurations than with other [BMB] conformers.
The energy gaps for bent B(R)-Man(RS) and B(S)-Man(SR) [BMB] conformers are slightly larger than those for twisted [BMB] configurations.
It is noteworthy that for the two V-shaped structures, the Li[BMB] (B(S)-Man(SS)) ion pair has a larger energy gap than that for Li[BMB] (B(R)-Man(RR)) ion pair.
This indicates that the Li[BMB] (B(S)-Man(SS)) ion pair has higher stability and opposing charge transfer than other Li[BMB] ion pair structures, since this ion pair complex opposes changes in its electron density and distribution.
In addition, the Li[BMB] (B(S)-Man(SS)) ion pair complex has a lower HOMO energy than that for Li[BMB] (B(R)-Man(RR)) ion pair structure, indicating that the former complex has a distinct electrochemical stability and a large electrochemical window that other Li[BMB] ion pair structures when it is used as solvent electrolyte in Li ion batteries~\cite{nilsson2019ionic, franco2019boosting}.

\par
\begin{figure}[!h]
\centering\includegraphics[width=0.8\textwidth]{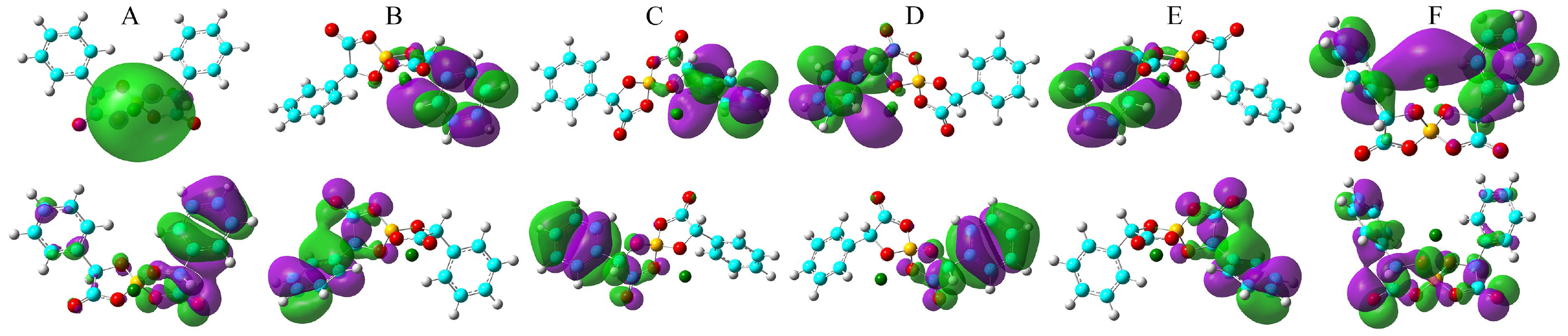}
\caption{The HOMO (lower panels) and LUMO (upper panels) contours for the optimized Li[BMB] ion pair structures with anions having varied molecular chiralities. These orbital isosurfaces were drawn with a contour value of 0.02. The purple and green isosurfaces of HOMO and LUMO indicates negative and positive values, respectively. The hydrogen, boron, carbon, and oxygen atoms in [BMB] anions are colored in white, yellow, cyan, and red beads, and Li$^+$ ions are represented by green spheres, respectively. (A) B(R)-Man(RR), (B) B(R)-Man(RS), (C) B(R)-Man(SS), (D) B(S)-Man(RR), (E) B(S)-Man(SR), and (F) B(S)-Man(SS).}\label{fig:libmb_homo_lumo_vis}
\end{figure}

The computational HOMO and LUMO contours for the optimized Li[BMB] ion pair structures are shown in Fig.~\ref{fig:libmb_homo_lumo_vis}.
It is clearly demonstrated that HOMO positions are mainly localized on phenyl groups and partially extended to oxalato ring structures for all Li[BMB] ion pair complexes.
This observation is attributed to the $\pi$-states of aromatic ring structures including both phenyl and oxalato ring moieties.
However, the LUMO distributions are concentrated over Li$^+$ ions for V-shaped B(S)-Man(SS) and B(R)-Man(RR) [BMB] conformers, and are located primarily around Li$^+$ ions and phenyl groups for twisted B(R)-Man(SS) and B(S)-Man(RR) [BMB] conformers, as well as for bent B(R)-Man(RS) and B(S)-Man(SR) [BMB] conformers.
The occurrence of electron transfer from HOMO to LUMO centers leads to an electron density transfer from one phenyl group to the other one in the same anion framework via central oxalato moieties.
Therefore, [BMB] anions are primarily responsible for the charge transfer process between Li$^+$ ions and anions.
Furthermore, we can assume that intramolecular charge transfer is enhanced within an anion framework for anions having larger ion hydrophobicity.

\par
\begin{figure}[!h]
\centering\includegraphics[width=0.8\textwidth]{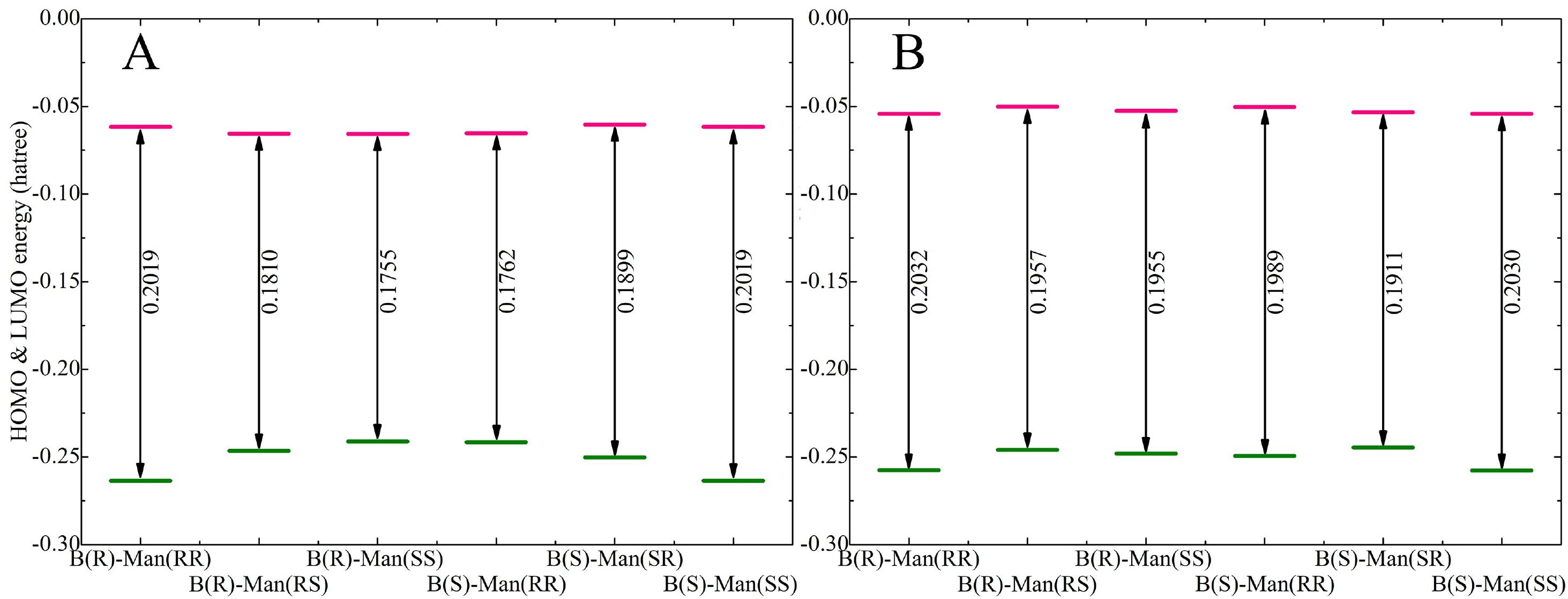}
\caption{Computational HOMO and LUMO energies (in Hartree), and energy gaps (LUMO-HOMO) for (A) Na[BMB] and (B) K[BMB] ion pair complexes with anions having varied molecular chiralities. These computational results were obtained from DFT calculations at B3LYP/6-311+G(d) level of theory with Grimme’s-D3 dispersion correction.}\label{fig:nakkbmb_homo_lumo_ene}
\end{figure}
\begin{figure}[!h]
\centering\includegraphics[width=0.8\textwidth]{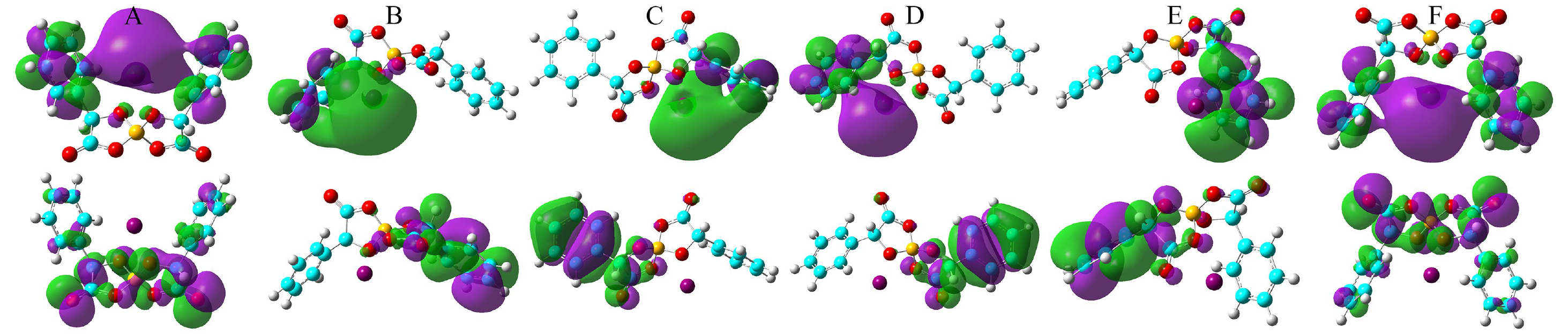}
\caption{The HOMO (lower panels) and LUMO (upper panels) contours for the optimized Na[BMB] ion pair structures with anions having varied molecular chiralities. These orbital isosurfaces were drawn with a contour value of 0.02. The purple and green isosurfaces of HOMO and LUMO indicates negative and positive values, respectively. The hydrogen, boron, carbon, and oxygen atoms in [BMB] anions are colored in white, yellow, cyan, and red beads, and Na$^+$ ions are represented by purple spheres, respectively. (A) B(R)-Man(RR), (B) B(R)-Man(RS), (C) B(R)-Man(SS), (D) B(S)-Man(RR), (E) B(S)-Man(SR), and (F) B(S)-Man(SS).}\label{fig:nabmb_homo_lumo_vis}
\end{figure}
\begin{figure}[!h]
\centering\includegraphics[width=0.8\textwidth]{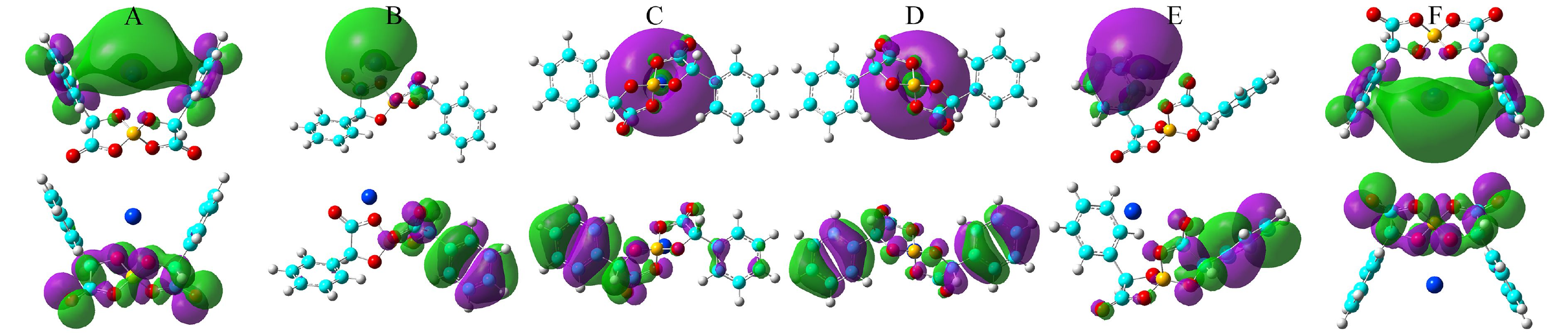}
\caption{The HOMO (lower panels) and LUMO (upper panels) contours for the optimized K[BMB] ion pair structures with anions having varied molecular chiralities. These orbital isosurfaces were drawn with a contour value of 0.02. The purple and green isosurfaces of HOMO and LUMO indicates negative and positive values, respectively. The hydrogen, boron, carbon, and oxygen atoms in [BMB] anions are colored in white, yellow, cyan, and red beads, and K$^+$ ions are represented by blue spheres, respectively. (A) B(R)-Man(RR), (B) B(R)-Man(RS), (C) B(R)-Man(SS), (D) B(S)-Man(RR), (E) B(S)-Man(SR), and (F) B(S)-Man(SS).}\label{fig:kkbmb_homo_lumo_vis}
\end{figure}

The HOMO and LUMO energy levels for the optimized Na[BMB] and K[BMB] ion pair structures are shown in Fig.~\ref{fig:nakkbmb_homo_lumo_ene}.
These HOMO and LUMO energies, and energy gaps exhibit similar features with those for Li[BMB] ion pair structures, indicating that these alkali metal ions have similar coordination patterns with [BMB] anions but with different interaction strength.
In addition, the HOMO and LUMO contours for the optimized Na[BMB] and K[BMB] ion pair structures shown in Fig.~\ref{fig:nabmb_homo_lumo_vis} and Fig.~\ref{fig:kkbmb_homo_lumo_vis} present a qualitative feature with those for the optimized Li[BMB] ion pair complexes shown in Fig.~\ref{fig:libmb_homo_lumo_vis} owing to a different ion parameters of Li$^+$, Na$^+$ and K$^+$ ions.
Nevertheless, the distinct coordination patterns and interaction strengths of different [BMB] conformers with three alkali metal ions may lead to varied stabilities and electrochemical windows of alkali metal ion-[BMB] complexes.
These striking physicochemical properties may contribute to striking applications of [BMB] anions in alkali metal ion batteries, in which alkali metal ion-[BMB] complexes are effective charge carriers contributing to significant ion conductivities in electrochemical devices.

% -------------------------------------------------------------------- %
\section{Atomistic Simulations of Tetraalkylphosphonium [BMB] Ionic Liquids}

\subsection{Atomistic simulation methodology}

\par
Atomistic force field parameters of tetraalkylphosphonium cations and [BMB] anions are taken from a systematically developed force field in our previous studies based on AMBER framework~\cite{wang2014atomistic, wang2015multiscale}.
The cross interaction parameters between different atom types are obtained from the Lorentz-Berthelot combination rules.
It is noteworthy that during force field development for tetraalkylphosphonium orthoborate ILs in our previous work~\cite{wang2014atomistic}, both tetraalkylphosphonium cations and orthoborate anions were described by unity charges and all atomic partial charges were determined by fitting molecular electrostatic potential generated from DFT calculations of individual ions.
In recent computational studies, it has been suggested that a down-scaling of atomic partial charges is an effective way to account for polarization and charge transfer effects between ion species, and thereby can improve reliability of calculations of dynamical and transport quantities of ILs~\cite{morrow2002molecular, ishizuka2016self}.
It was identified in a recent study that a charge scaling factor of $0.8$ could provide reliable liquid viscosities of [P$_{6,6,6,14}$][BMB] and [P$_{6,6,6,14}$]Cl ILs as those obtained from experimental measurements within a wide temperature interval of $373$-$463$ K~\cite{sarman2018rheology}.
Therefore in current work, all atomic partial charges for atoms in tetraalkylphosphonium cations and [BMB] anions are uniformly rescaled using a scaling factor of $0.8$.
It should be noted that even the adopted charge scaling factor of $0.8$ is an empirical correction of electrostatic polarization and charge transfer effects between tetraalkylphosphonium cations and [BMB] anions, it gives a good performance in describing thermodynamics and microstructural quantities of tetraalkylphosphonium [BMB] ILs.

\par
In present atomistic simulations, each simulation system consists of varied number of tetraalkylphosphonium [BMB] ion pairs having approximately $35000$ atoms.
More specifically, [P$_{\textrm{4,4,4,4}}$][BMB], [P$_{\textrm{4,4,4,8}}$][BMB], and [P$_{\textrm{6,6,6,14}}$][BMB] IL systems are composed of $400$, $352$, and $259$ ion pairs, respectively.
Atomistic simulations were performed using GROMACS 5.0.7 package~\cite{abraham2015gromacs} with cubic periodic boundary conditions.
The equations of motion were integrated using a classical velocity Verlet leapfrog integration algorithm with a time step of $1.0$ fs.
A cutoff distance of $1.6$ nm was set for short-range van der Waals interactions and real-space electrostatic interactions between atomic partial charges.
The Particle-Mesh Ewald summation method with an interpolation order of $5$ and a Fourier grid spacing of $0.16$ nm was employed to handle long range electrostatic interactions in reciprocal space.

\par
All tetraalkylphosphonium [BMB] IL systems were first energetically minimized using a steepest descent algorithm, and thereafter annealed gradually from $800$ to $323$ K within $20$ ns.
All annealed simulation systems were equilibrated in NPT (isothermal-isobaric) ensemble for $60$ ns of physical time maintained using Nos\'e-Hoover chain thermostat and Parrinello-Rahman barostat with time coupling constants of $0.5$ and $0.2$ ps, respectively, to control temperature at $323$ K and pressure at $1$ atm.
Atomistic simulations of all simulation systems were further performed in NVT ensemble for $100$ ns, and simulation trajectories were recorded at an interval of $100$ fs for further microstructural analysis.
Additional atomistic simulations were performed at a wide temperature range to address the dependence of liquid densities of all studied tetraalkylphosphonium [BMB] ILs on temperatures for a comparative purpose.

\subsection{Liquid densities}

\par
\begin{figure}[!h]
\centering\includegraphics[width=0.8\textwidth]{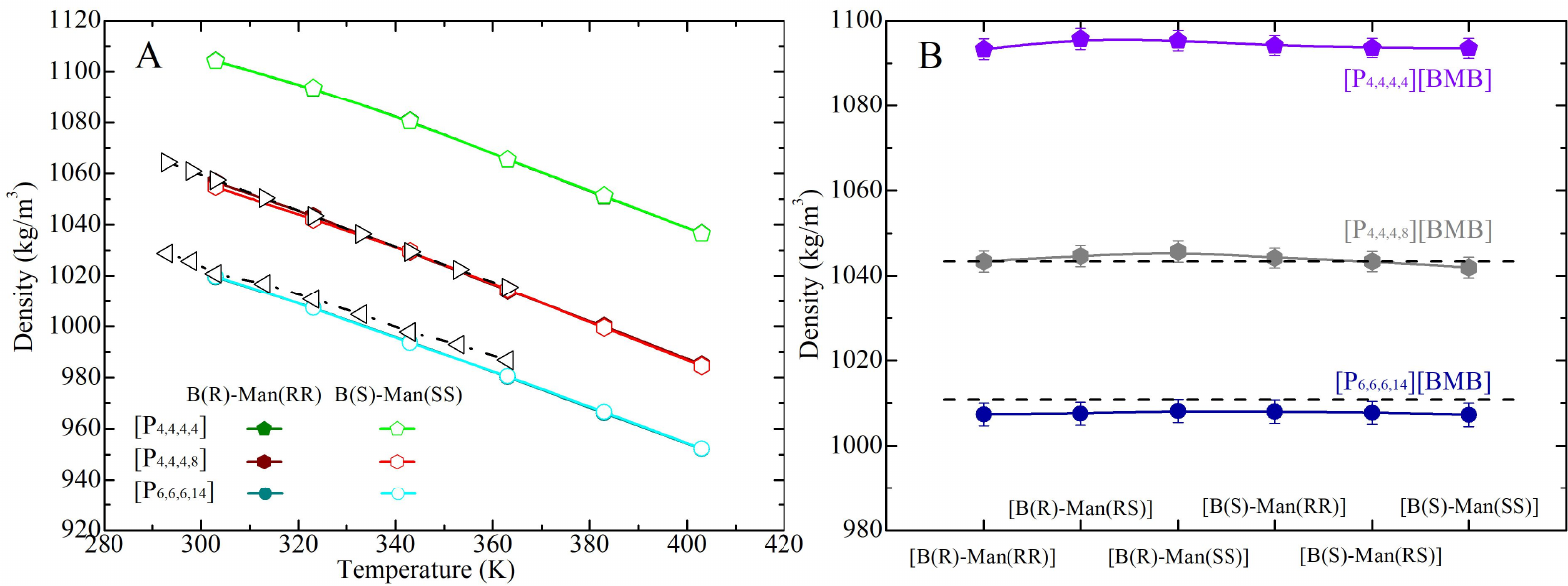}
\caption{Liquid densities of (A) representative tetraalkylphosphonium [BMB] ILs at different temperatures and (B) all studied tetraalkylphosphonium [BMB] ILs at $323$ K. Experimental data (dashed lines) for $[\textrm{P}_{4,4,4,8}]$[BMB] and $[\textrm{P}_{6,6,6,14}]$[BMB] ILs are derived from Ref.~\cite{shah2011novel} for comparison.}\label{fig:density}
\end{figure}

Liquid densities of all studied tetraalkylphosphonium [BMB] IL systems are calculated from current atomistic simulations at different temperatures ranging from $293$ to $373$ K with an interval of $20$ K, and representative computational results are shown in Fig.~\ref{fig:density}.
In addition, experimental density data of $[\textrm{P}_{4,4,4,8}]$[BMB] and $[\textrm{P}_{6,6,6,14}]$[BMB] ILs taken from Ref.~\cite{shah2011novel} are also provided in Fig.~\ref{fig:density} for a comparative propose.
The liquid densities of all tetraalkylphosphonium [BMB] ILs exhibit linear variations as temperature changes within the investigated temperature range.
For a given [BMB] conformer, both experimental and computational density results of tetraalkylphosphonium [BMB] ILs decrease with increasing the number of carbon atoms in cations at specific temperatures following an order of $[\textrm{P}_{4,4,4,4}]$ $>$ $[\textrm{P}_{4,4,4,8}]$ $>$ $[\textrm{P}_{6,6,6,14}]$.
This is attributed to the reduced interactions between large tetraalkylphosphonium cations and [BMB] anions leading to their less efficient packing in heterogeneous IL matrices~\cite{wang2014atomistic, wang2017microstructures}.
In addition, a good agreement is observed between experimental data and computational results for $[\textrm{P}_{4,4,4,8}]$[BMB] and $[\textrm{P}_{6,6,6,14}]$[BMB] ILs at all studied temperatures.

\par
However, molecular chiralities of [BMB] anions have negligible effect on liquid densities of these tetraalkylphosphonium [BMB] ILs, which may attribute to elaborate interactions among all constituent ions in liquid environments.
It is known that liquid density is a thermodynamic manifestation of all possible ion pair and ion cluster structures in local liquid environments.
Even [BMB] conformers have their intrinsic molecular volumes as determined from DFT calculations, these [BMB] anions have delicate coordinations with tetraalkylphosphonium cations in heterogeneous IL matrices, leading to their comparable liquid densities as shown in Fig.~\ref{fig:density}(B).
In addition, the agreement between experimental data and simulation results is remarkably good with a maximum deviation of approximately 0.3\% and 0.5\% for $[\textrm{P}_{4,4,4,8}]$[BMB] and $[\textrm{P}_{6,6,6,14}]$[BMB] ILs, respectively.
These computational data suggest that there might be multiple [BMB] conformers in experimental tetraalkylphosphonium [BMB] IL samples depending on the relative ratio of R- and S-mandelic acids in reaction vessel.
Therefore, advanced synthesis method and separation technology will be useful for a better extraction of varied [BMB] conformers from IL samples due to their specific coordinations with cation groups, like alkali metal ions.

\subsection{Hydrogen bonding interactions}

\par
HB interactions between ion species in ILs are one of the most important interactions, and have a significant effect on ion packing structures in local ionic environments~\cite{matthews2015hydrogen, hunt2015hydrogen, sha2016origin, wang2017hydrogen, wang2018competitive, pei2019feature, wang_pxxxxntf2}.
In current work, we studied HB interactions between tetraalkylphosphonium cations and [BMB] anions having varied molecular chiralities.
For all these tetraalkylphosphonium [BMB] ILs, HP atoms in cations (as labelled in Fig.~\ref{fig:il_structure}) are potential HB donor sites~\cite{wang2017microstructures, wang_pxxxxntf2} and all three oxygen atom types in anions (as labelled in Fig.~\ref{fig:il_structure}) are preferential HB acceptors, respectively.

\begin{figure}[!h]
\centering\includegraphics[width=0.8\textwidth]{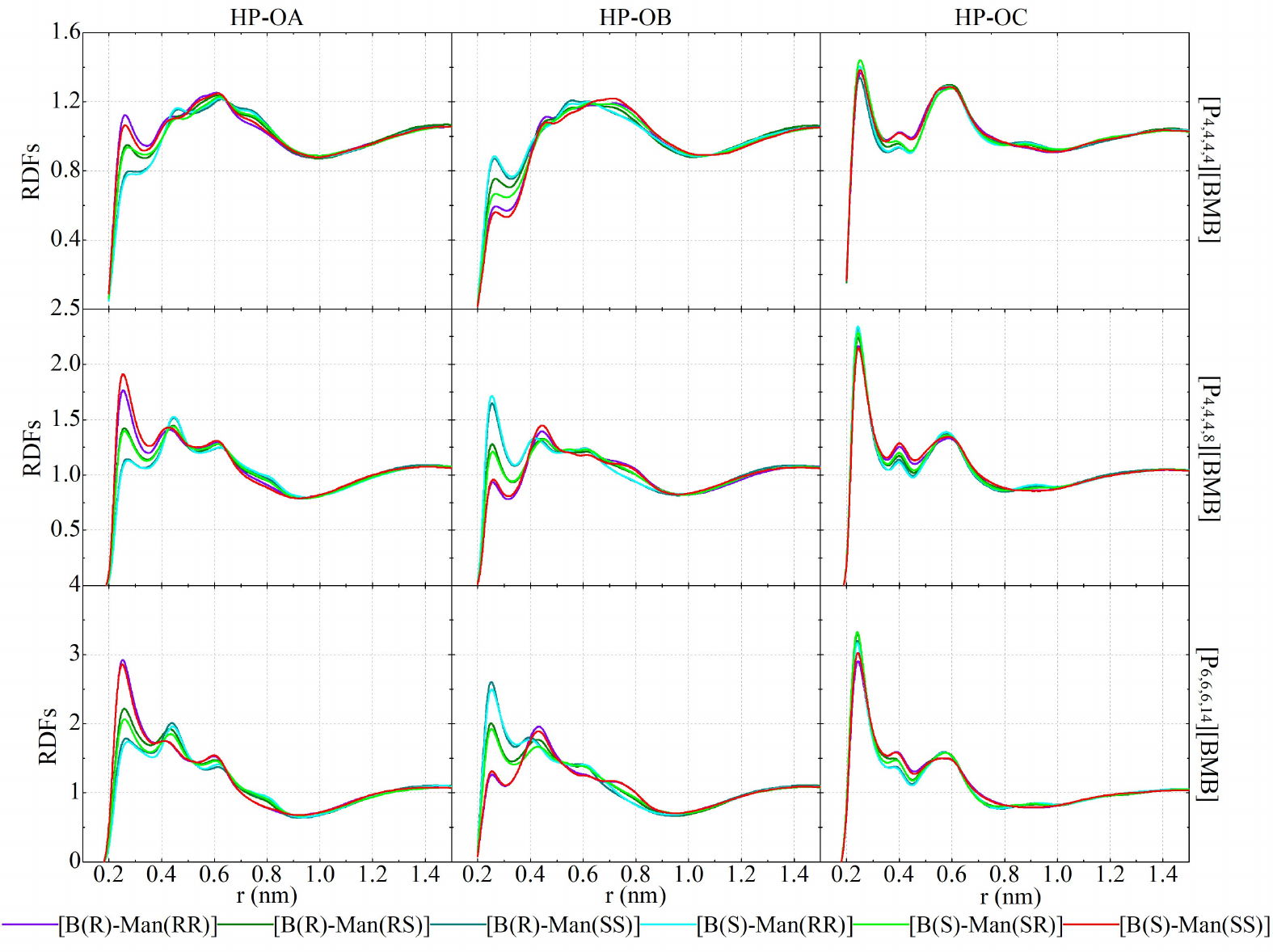}
\caption{RDFs of HP atoms in [P$_{4,4,4,4}$] (upper panels), [P$_{4,4,4,8}$] (middle panels), and [P$_{6,6,6,14}$] (lower panels) cations with three oxygen atom types (OA, OB, and OC atoms as labelled in Fig.~\ref{fig:il_structure}) in [BMB] anions having varied molecular chiralities.}\label{fig:hbrdf}
\end{figure}

\par
Fig.~\ref{fig:hbrdf} presents representative HP-O radial distribution functions (RDFs) for HB coordinations of tetraalkylphosphonium cations with [BMB] anions.
For given tetraalkylphosphonium cations, the RDF peak intensities of HP atoms in coordinating OA and OB atoms in [BMB] anions follow three categories of [BMB] conformers as discussed in previous section.
The V-shaped B(R)-Man(RR) and B(S)-Man(SS) [BMB] conformers have strong HB associations, the bent B(R)-Man(RS) and B(S)-Man(SR) [BMB] conformers have intermediate HB interactions, and the twisted B(R)-Man(SS) and B(S)-Man(RR) [BMB] conformers have weak HB coordinations with HP atoms in tetraalkylphosphonium cations via OA atoms (left column in Fig.~\ref{fig:hbrdf}), respectively.
However, an opposite effect is observed for HP atoms in HB interactions with OB atoms in [BMB] anions, as clearly manifested in middle column in Fig.~\ref{fig:hbrdf}.
In addition, OC atoms in all [BMB] conformers have stronger interactions with HP atoms in tetraalkylphosphonium cations than OA and OB atoms, irrespective of molecular chiralities of [BMB] conformers, as shown in the right column in Fig.~\ref{fig:hbrdf}.
These computational results indicate that OA and OB have competitive HB interactions with HP atoms in tetraalkylphosphonium cations, and their HB capabilities depend on molecular chiralities, conformational flexibilities, and steric hinderance effects of [BMB] conformers.
The V-shaped B(R)-Man(RR) and B(S)-Man(SS) [BMB] conformers have small cavities in phenyl-oxalato-phenyl domains, which prevent central polar groups of tetraalkylphosphonium cations from approaching, and thus OB atoms in local cavities have weak HP-OB interactions.
The HB acceptor sites outside cavities, such as OC and OA atoms, have preferential HB interactions with HP atoms in tetraalkylphosphonium cations.
For bent B(R)-Man(RS) and B(S)-Man(SR) [BMB] conformers, both two OA and two OB atoms have one site inside and one outside local cavities, and thus OA and OB atoms have comparable HB associations with neighboring tetraalkylphosphonium cations via HP atoms.
The OA and OB atoms have opposite distributions in twisted B(R)-Man(SS) and B(S)-Man(RR) [BMB] conformers as those in V-shaped B(R)-Man(RR) and B(S)-Man(SS) [BMB] conformers, and thus they have distinct HB coordinations with tetraalkylphosphonium cations via HP atoms.

\par
Enlarging tetraalkylphosphonium cation sizes from [P$_{4,4,4,4}$] (upper panels in Fig.~\ref{fig:hbrdf}) to [P$_{4,4,4,8}$] (middle panels in Fig.~\ref{fig:hbrdf}) and [P$_{6,6,6,14}$] (lower panels in Fig.~\ref{fig:hbrdf}) leads to enhanced HB interactions of HP in cations with all three oxygen atom types in [BMB] anions.
This observation is rationalized by enhanced segregation of polar groups consisting of central oxalato moieties in anions and central P(CH$_2$)$_4$ groups in cations in apolar networks consisting of phenyl groups in anions and remaining alkyl moieties in cations with a gradual addition of apolar alkyl groups to tetraalkylphosphonium cations, as has been intensively discussed in previous works~\cite{wang2017microstructures, wang2018competitive}.

\subsection{Microscopic liquid structures}

\par
\begin{figure}[!h]
\centering\includegraphics[width=0.8\textwidth]{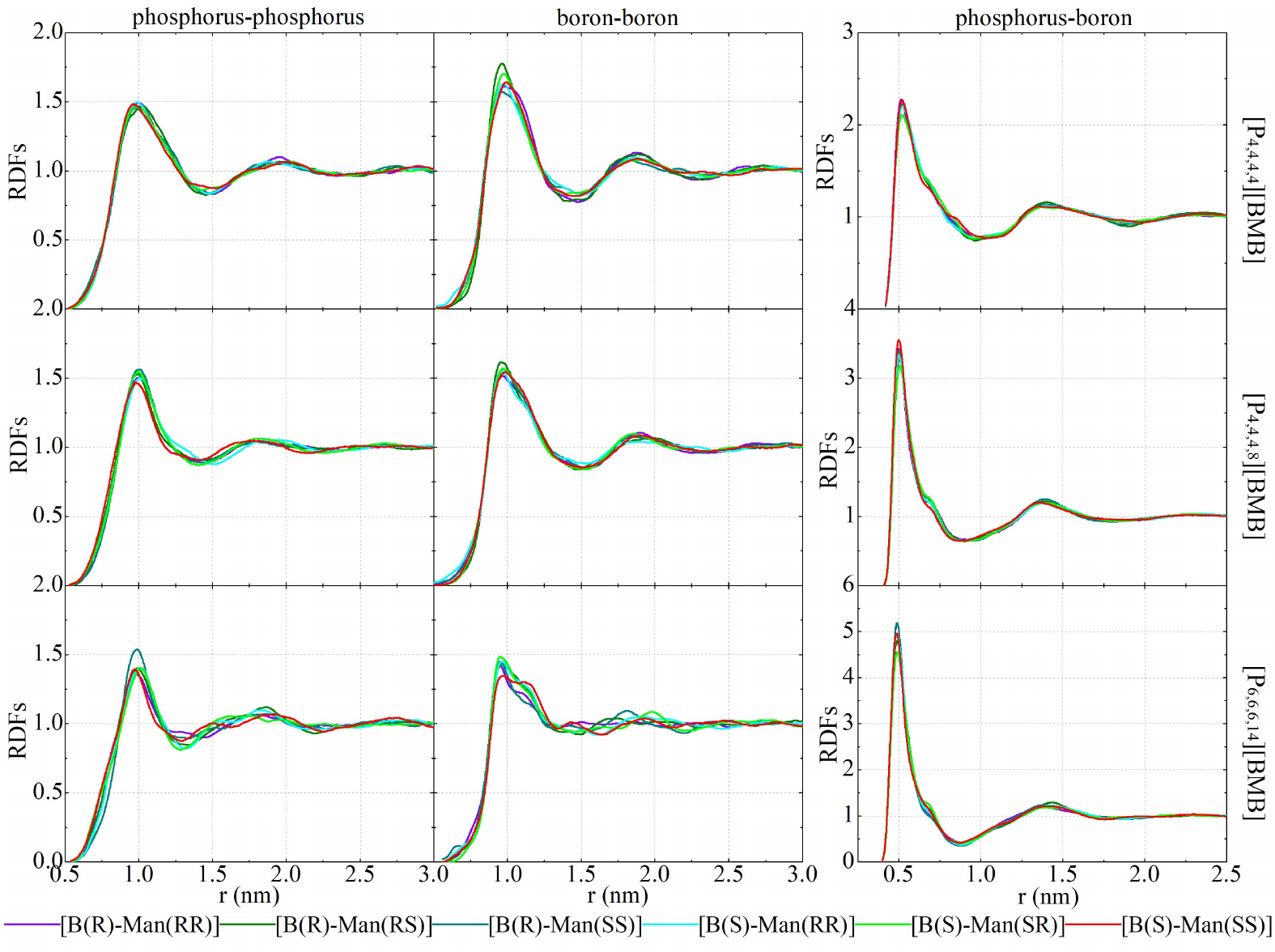}
\caption{RDFs between phosphorus atoms in cations and boron atom in anions in [P$_{4,4,4,4}$][BMB] (upper panels), [P$_{4,4,4,8}$][BMB] (middle panels), and [P$_{6,6,6,14}$][BMB] (lower panels) ILs with anions having varied molecular chiralities.}\label{fig:comrdf}
\end{figure}

Fig.~\ref{fig:comrdf} presents representative site-site RDFs of cation-cation, anion-anion, and cation-anion pairs for all studied tetraalkylphosphonium [BMB] ILs.
The central phosphorus atoms in tetraalkylphosphonium cations and boron atoms in [BMB] anions are taken as reference sites to calculate these RDFs, which are useful to depict microscopic liquid structures between ion species in tetraalkylphosphonium [BMB] ILs.
Both phosphorus-phosphorus and boron-boron RDF plots present very broad peaks at around $1.0$ nm with similar peak intensities in all studied tetraalkylphosphonium [BMB] ILs.
Molecular chiralities of [BMB] anions have minimal effect on spatial distributions of boron atoms in [BMB] anions around phosphorus atoms in tetraalkylphosphonium cations, and~\emph{vice versa}.
These computational data are rationalized by strong Coulombic interactions among phosphorus and boron atoms in tetraalkylphosphonium cations and [BMB] anions, respectively.
In addition, Coulombic interactions between central polar groups in tetraalkylphosphonium cations and [BMB] anions are overwhelmingly stronger than preferential intermolecular HB interactions, the latter of which are difficult to mediate distributions of anions around cations, but to fine-tune local orientations of [BMB] conformers to maximize their HB interactions with tetraalkylphosphonium cations in local ionic environments.

\par
Enlarging molecular sizes of tetraalkylphosphonium cations from [P$_{4,4,4,4}$] (upper panels in Fig.~\ref{fig:comrdf}) to [P$_{4,4,4,8}$] (middle panels in Fig.~\ref{fig:comrdf}) and [P$_{6,6,6,14}$] (lower panels in Fig.~\ref{fig:comrdf}) leads to increased intermolecular interactions between phosphorus and boron atoms in respective ion groups, as clearly manifested in right column in Fig.~\ref{fig:comrdf}, owing to enhanced segregation of polar groups in apolar networks in heterogeneous IL matrices~\cite{wang2017microstructures, wang2018competitive}.

\par
In addition, we calculated the total X-ray scattering structural function, $S(q)$, using a summation of atom type based partial components as $S(q)=\sum_{i=1}^n \sum_{j=1}^n S_{ij}(q)$, to explore an overall effect of molecular chiralities of [BMB] anions and variations of tetraalkylphosphonium cation structures on microstructural ordering characteristics in tetraalkylphosphonium [BMB] ILs.
The $S_{ij}(q)$ is a partial structural function between atoms types $i$ and $j$ and is given by
\begin{eqnarray}
S_{ij}(q)=\frac{\rho_0 x_i x_j f_i(q) f_j(q) \int_0^{L/2} 4\pi r^2 [g_{ij}(r)-1] \frac{sin(qr)}{qr}W(r)dr}{[\sum_{i=1}^n x_i f_i(q)]^2}\,.\nonumber
\end{eqnarray}
$x_i$ and $x_j$ are mole fractions, and $f_i(q)$ and $f_j(q)$ are X-ray atomic form factors of atom types $i$ and $j$ in simulation systems~\cite{prince2004international}, respectively.
$g_{ij}(r)$ is the partial RDF between atom types $i$ and $j$, including both intra- and intermolecular pairs.
$\rho_0=\frac{N_{atom}}{<L^3>}$ refers to an averaged atom number density of a simulation system and $L$ is the simulation box length.
$W(r)$ is a Lorch window function defined as $W(r)=\frac{sin(2\pi r/L)}{2\pi r/L}$, which is used to minimize effect of finite truncation of $r$ for the calculation of $g_{ij}(r)$.
For a simulation system of moderate size, this window function does not hinder the physical meaning of peaks and anti-peaks in $S(q)$ plots~\cite{annapureddy2010origin, hettige2012anions, wu2016structure, wang2017microstructures}.

\begin{figure}[!h]
\centering\includegraphics[width=0.8\textwidth]{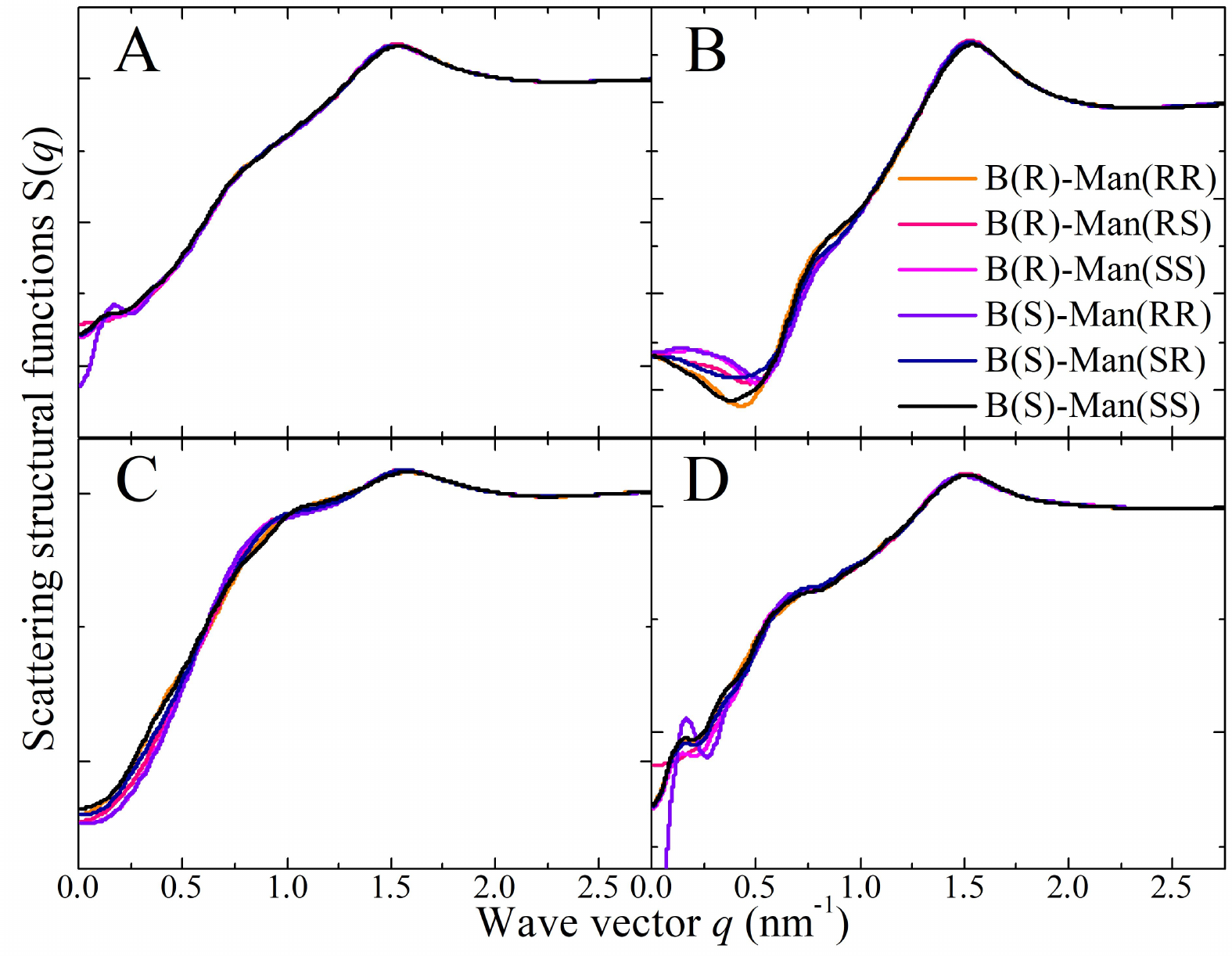}
\caption{(A) The total X-ray scattering structural functions $S(q)$ in the range of $q \leq 2.75$ nm$^{-1}$ for [P$_{4,4,4,4}$][BMB] IL at $323$ K with anions having varied molecular chiralities. These total X-ray scattering structural functions are further partitioned into partial structural functions for (B) cation-anion/anion-cation, (C) cation-cation, and (D) anion-anion subcomponents.}\label{fig:4444_sq_tt_ca_an}
\end{figure}

\par
Fig.~\ref{fig:4444_sq_tt_ca_an}A presents the total X-ray scattering structural functions $S(q)$ for [P$_{4,4,4,4}$][BMB] ILs in the range of $q \leq 2.75$ nm$^{-1}$, which is the most relevant region and is associated with intermolecular correlations between tetraalkylphosphonium cations and [BMB] anions in heterogeneous IL matrices.
Features at $q$ values larger than 2.75 nm$^{-1}$ are mostly intramolecular in nature, which are fairly easy to assign and thus are not discussed in current work.
Two prominent peaks located at $\sim$0.76 nm$^{-1}$ and $\sim$1.51 nm$^{-1}$ are shown in these total X-ray scattering structural functions for [P$_{4,4,4,4}$][BMB] ILs with anions having varied molecular chiralities.
Such a two-peak-plot is a general feature for tetraalkylphosphonium ILs, and is mainly attributed to bulky and voluminous structures of cations and their strong coordinations with anions via strong Coulombic interactions between polar groups in respective ion species and preferential hydrophobic interactions among alkyl and phenyl moieties in constituent ions~\cite{amith2016structures, hettige2016nanoscale, wang2017microstructures, wang_pxxxxntf2}.
These peak positions are essential characteristic hallmarks of their microstructural landscapes, indicating particular microscopic ion ordering phenomena at different length scales in tetraalkylphosphonium [BMB] IL matrices~\cite{annapureddy2010origin, hettige2012anions, amith2016structures, wu2016structure, wang2017microstructures, wang2018competitive, wang_pxxxxntf2}.

\par
The peaks at intermediate $q$ range around 0.76 nm$^{-1}$ are an indicative of mesoscopic liquid organization characterized by positive-negative charge alternations in IL matrices~\cite{canongia2006nanostructural, annapureddy2010origin, amith2016structures, wu2016structure, wang2017microstructures, weyman2018microphase, wang_pxxxxntf2}.
This charge ordering behavior is the need to maintain a lattice-like arrangement of cations and anions to minimize Coulombic interactions in IL matrices, and thus is associated with a length scale between ions of the same charge separated by ions of opposite charge~\cite{annapureddy2010origin, hettige2012anions, wu2016structure, wang2017microstructures, wang2018competitive, weyman2018microphase, wang_pxxxxntf2}.
For some ILs, this charge alternation peak is present as a weak shoulder because of an almost complete cancellation of peaks and anti-peaks that offset this ordering phenomena at intermediate length scale~\cite{amith2016structures, wang2017microstructures}.
The prominent peaks at high $q$ values near 1.51 nm$^{-1}$ are associated with short range adjacency correlations originated from nearest neighboring ion species~\cite{annapureddy2010origin, hettige2012anions, hettige2016nanoscale, wu2016structure, wang2017microstructures, wang2018competitive, wang_pxxxxntf2}.
In fact, this peak mainly originates from apolar adjacency correlations between ion species, as has been verified in previous studies~\cite{hettige2012anions, amith2016structures, wu2016structure, wang2017microstructures, weyman2018microphase}.

\par
Either for the total X-ray scattering structural functions or the partial structural functions for cation-anion/anion-cation (Fig.~\ref{fig:4444_sq_tt_ca_an}B), cation-cation (Fig.~\ref{fig:4444_sq_tt_ca_an}C), and anion-anion (Fig.~\ref{fig:4444_sq_tt_ca_an}D) subcomponents in [P$_{4,4,4,4}$][BMB] ILs, the dependence of charge alternations and adjacency correlations in these scattering structural functions on molecular chiralities of [BMB] anions is negligible.
This is expected as scattering structural functions are intrinsically determined by RDFs between different atom pairs.
It is shown in Fig.~\ref{fig:comrdf} that molecular chiralities of [BMB] anions have minimal effects on phosphorus-phosphorus, phosphorus-boron, and boron-boron RDFs, which leads to comparable scattering structural functions for [P$_{4,4,4,4}$][BMB] ILs with anions have varied molecular chiralities.
Therefore, it can be addressed that [P$_{4,4,4,4}$][BMB] ILs are characterized by similar liquid morphologies, which are independent of molecular chiralities of [BMB] anions.
Additional computational results addressing that molecular chiralities of [BMB] anions have minimal influence on the total and partial scattering structural functions were also obtained for [P$_{4,4,4,8}$][BMB] and [P$_{6,6,6,14}$][BMB] ILs.

\begin{figure}[!h]
\centering\includegraphics[width=0.8\textwidth]{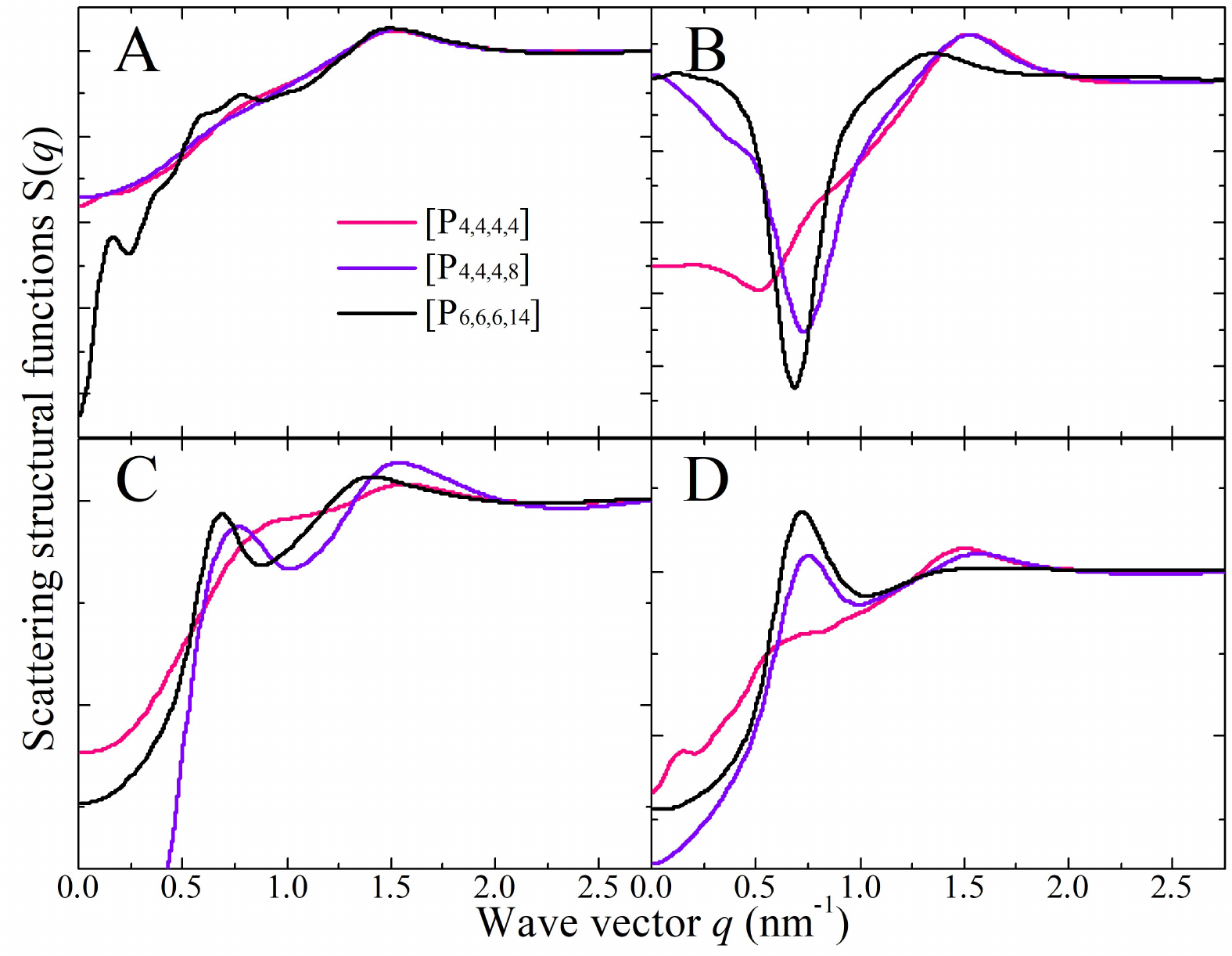}
\caption{(A) The total X-ray scattering structural functions $S(q)$ in the range of $q \leq 2.75$ nm$^{-1}$ for tetraalkylphosphonium [BMB] ILs at $323$ K with anions characterized by B(R)-Man(SS) configuration. These total X-ray scattering structural functions are further partitioned into partial structural functions for (B) cation-anion/anion-cation, (C) cation-cation, and (D) anion-anion subcomponents.}\label{fig:bmb3_sq_tt_ca_an}
\end{figure}

\par
Fig.~\ref{fig:bmb3_sq_tt_ca_an} presents representative total and partial X-ray scattering structural functions for tetraalkylphosphonium [BMB] ILs at $323$ K with anions characterized by B(R)-Man(SS) configuration.
In intermediate $q$ range, both cation-cation (Fig.~\ref{fig:bmb3_sq_tt_ca_an}C) and anion-anion (Fig.~\ref{fig:bmb3_sq_tt_ca_an}D) subcomponents contribute to positive intensities, and the cross term cation-anion/anion-cation subcomponent (Fig.~\ref{fig:bmb3_sq_tt_ca_an}B) exhibits prominent anti-peaks at the same $q$ range, respectively.
A combination of these peaks and anti-peaks contributes to prominent peaks registered at intermediate $q$ range in the total X-ray scattering structural functions shown in Fig.~\ref{fig:bmb3_sq_tt_ca_an}A.
Independent of tetraalkylphosphonium [BMB] ILs, the charge alternation peaks always appear as a result of cancellation of prominent positive contributions from same charge subcomponents and distinct negative-going peaks from cross term contributions~\cite{annapureddy2010origin, amith2016structures, wu2016structure, wang2017microstructures, wang2018competitive, weyman2018microphase}.
This observation indicates that at specific locations where one expects to find a same-charge ion there is a systematic absence of ions having opposite charge~\cite{annapureddy2010origin, hettige2012anions, amith2016structures, hettige2016nanoscale, wu2016structure, wang2017microstructures, wang2018competitive, wang_pxxxxntf2}.
In addition, all partial scattering structural functions present positive contributions to adjacency correlations between neighboring ion species at high $q$ values around 1.51 nm$^{-1}$ but with different portions of contribution.

\par
\begin{figure}[!h]
\centering\includegraphics[width=0.8\textwidth]{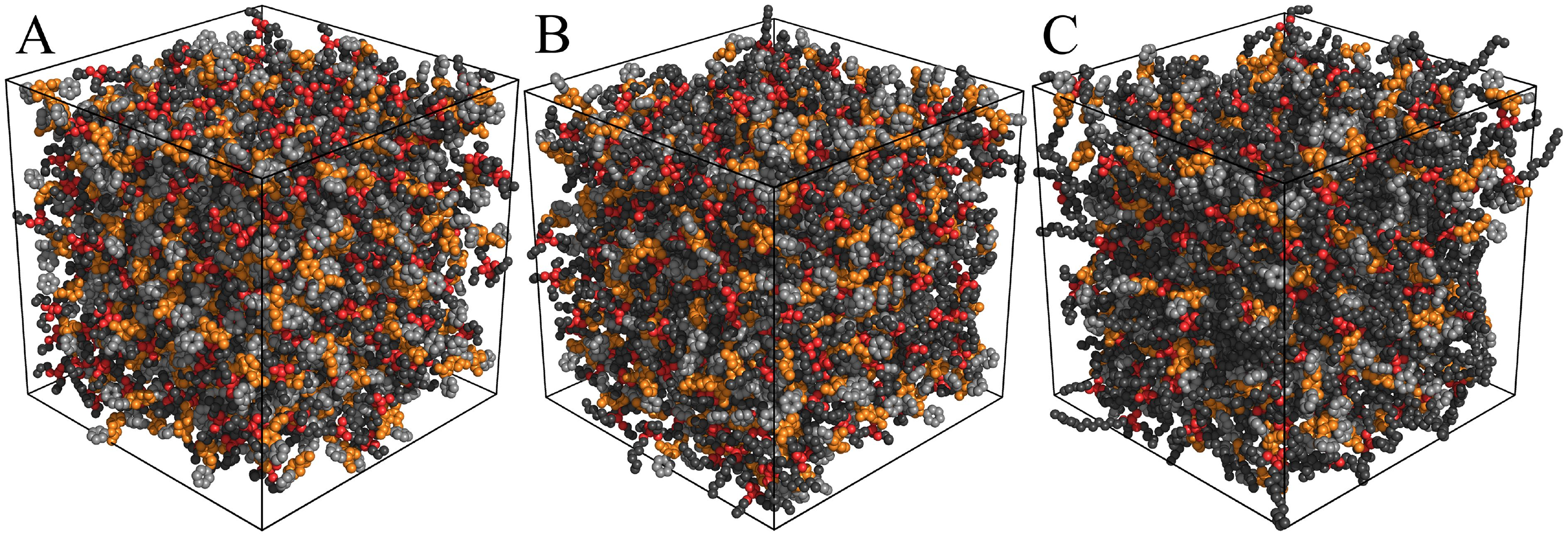}
\caption{Representative liquid morphologies of (A) [P$_{4,4,4,4}$][BMB], (B) [P$_{4,4,4,8}$][BMB], and (C) [P$_{6,6,6,14}$][BMB] ILs with anions characterized by B(R)-Man(SS) configuration. Polar domains consist of oxalato moieties (orange) in [BMB] anions and central P(CH$_2$)$_4$ groups (red) in tetraalkylphosphonium cations, and apolar entity is composed of phenyl units (light grey) in [BMB] anions and the remaining alkyl groups (dark grey) in tetraalkylphosphonium cations.}\label{fig:bmb3_morphology}
\end{figure}

Different to the effect of molecular chiralities of [BMB] anions on scattering structural functions for tetraalkylphosphonium [BMB] ILs, the variation of cation structures from [P$_{4,4,4,4}$] to [P$_{4,4,4,8}$] and [P$_{6,6,6,14}$] has a significant effect on RDFs (Fig.~\ref{fig:comrdf}), scattering structural functions (Fig.~\ref{fig:bmb3_sq_tt_ca_an}), and microscopic liquid morphologies (Fig.~\ref{fig:bmb3_morphology}).
Enlarging cation sizes from [P$_{4,4,4,8}$] to [P$_{6,6,6,14}$] leads to a concomitant shift of scattering peaks for charge alternations and adjacency correlations to lower $q$ values (corresponding to larger characteristic distances for apolar domains in real space) in total and partial scattering structural functions~\cite{amith2016structures, wu2016structure, wang2017microstructures, wang2018competitive, wang_pxxxxntf2}.
These computational results are rationalized by dispersed distributions of polar domains consisting of central oxalato moieties in anions and central P(CH$_2$)$_4$ groups in cations in apolar networks consisting of phenyl groups in anions and remaining alkyl moieties in cations, and the expansion of apolar networks in IL matrices with lengthening alkyl chains in tetraalkylphosphonium cations, as shown in Fig.~\ref{fig:bmb3_morphology}.
Such a microstructural change in tetraalkylphosphonium [BMB] IL matrices has an impact on the number of counterions present in the solvation shells of a given ion, and an even bigger effect on the distributions of (same-charge) ions nearby, and thus contributes to distinct changes in peak positions for charge alternations and adjacency correlations at low and high $q$ values, respectively.
This observation is, in general, consist with computational results for imidazolium oxalatoborate ILs with cations having varied alkyl chains~\cite{wang2017hydrogen, wang2018competitive} and for ILs consisting of tetraalkylphosphonium cations coupled with chloride, bromide, dicyanamide, and bis(trifluoromethylsulfonyl)imide anions~\cite{hettige2012anions, hettige2016nanoscale, wang2017microstructures, wang_pxxxxntf2}.

% -------------------------------------------------------------------- %
\section{Concluding Remarks}

\par
Spiroborate anions based inorganic electrolytes and ILs have fascinating physicochemical and structural properties, and have promising applications in tribology and electrochemistry.
Besides variations in cation structures, molecular chiralities of spiroborate anions have a significant effect on their macroscopic functionalities in these applications due to their specific binding structures and preferential interactions with neighboring ions in liquid matrices.
A thorough understanding of delicate associations between spiroborate anions and the paired cation species will provide valuable information for the selection and design of suitable lubricants and solvent electrolytes with desirable physicochemical properties for their specific applications.

\par
In current study, we performed intensive DFT calculations to study specific binding structures of [BMB] anions having varied molecular chiralities with representative alkali metal ions and representative electronic properties of alkali metal ion-[BMB] ion pair complexes.
The optimized [BMB] conformers are characterized by V-shaped, bent, and twisted anion structures with varied electrostatic potential contours, conformational energies, and distinct alkali metal ion-[BMB] binding structures.
Alkali metal ions have considerable associations with phenyl groups in V-shaped [BMB] conformers owing to preferential cation-$\pi$ interactions.

\par
In addition, we carried out extensive atomistic interactions to explore effects of molecular chiralities of [BMB] anions on thermodynamics and microstructural properties of bulk tetraalkylphosphonium [BMB] ILs.
It was revealed that oxygen atoms in [BMB] anions have competitive hydrogen bonding interactions with hydrogen atoms in tetraalkylphosphonium cations depending on molecular chiralities of [BMB] anions and steric hindrance effects of phenyl groups in [BMB] anions.
However, molecular chiralities of [BMB] anions have little effect on liquid densities of tetraalkylphosphonium [BMB] ILs and spatial distributions of boron atoms in anions around phosphorous atoms in cations.
Enlarging tetraalkylphosphonium cation sizes from [P$_{4,4,4,4}$] to [P$_{4,4,4,8}$] and [P$_{6,6,6,14}$] leads to enhanced cation-anion intermolecular hydrogen bonding and Coulombic interactions, as well as distinct microscopic liquid structures characterized by computational X-ray scattering structural functions, due to enhanced segregation of polar groups (consisting of central oxalato moieties in [BMB] anions and central P(CH$_2$)$_4$ groups in tetraalkylphosphonium cations) in apolar networks (composed of phenyl groups in [BMB] anions and the remaining alkyl moieties in tetraalkylphosphonium cations) in heterogeneous IL matrices.

% -------------------------------------------------------------------- %
%\clearpage
\section*{Acknowledgment}
Y.-L. Wang gratefully acknowledges financial support from Knut and Alice Wallenberg Foundation.
A. Laaksonen acknowledges Swedish Science Council for financial support, and thanks for partial support by a grant from Ministry of Research and Innovation, CNCS-UEFISCDI, project number PN-III-P4-ID-PCCF-2016-0050, within PNCDI III.
All molecular simulations were performed using computational resources provided by Swedish National Infrastructure for Computing (SNIC) at PDC, HPC2N and NSC.

% -------------------------------------------------------------------- %
\bibliography{bmb_chirality}

\end{document}